\documentclass[aps,twocolumn,prb,showpacs,10pt,floatfix,groupedaddress]{revtex4-1}
\usepackage{amssymb}
\usepackage{amsmath}
\usepackage{graphicx}
\usepackage{braket}
\usepackage{dcolumn}
\usepackage{bm}
\usepackage{textcomp}
\usepackage{color}
\usepackage{braket}
\usepackage{amsthm}
\usepackage{url}
\usepackage{mathtools}

\begin{document}

\title{Topological flat bands in time-periodically driven uniaxial strained graphene nanoribbons}

\author{Pedro Roman-Taboada}  
\email{peter89@fisica.unam.mx}
\author{Gerardo G. Naumis}

\affiliation{Departamento de Sistemas Complejos, Instituto de
F\'{i}sica, Universidad Nacional Aut\'{o}noma de M\'{e}xico (UNAM),
Apartado Postal 20-364, 01000 M\'{e}xico, Distrito Federal,
M\'{e}xico}


\begin{abstract}
We study the emergence of electronic non-trivial topological flat bands in time-periodically driven strained graphene within a tight binding approach based on the Floquet formalism. In particular, we focus on uniaxial spatially periodic strain since it can be mapped onto an effective one-dimensional system. Also, two kinds of time-periodic driving are considered: 
a short pulse (delta kicking) and a sinusoidal variation (harmonic driving). We prove that for special strain wavelengths, the system is described by a two level Dirac Hamiltonian. Even though the study case is gapless, we find that topologically non-trivial flat bands emerge not only at zero-quasienergy but also at $\pm\pi$ quasienergy, the latter being a direct consequence of the periodicity of the Floquet space. Both kind of flat bands are thus understood as dispersionless bands joining two inequivalent touching band points with opposite Berry phase. This is confirmed by explicit evaluation of the Berry phase in the touching band points' neighborhood. Using that information, the topological phase diagram of the system is built. Additionally, the experimental feasibility of the model is discussed and two methods for the experimental realization of our model are proposed.
\end{abstract}


\maketitle

\section{Introduction}
\label{intro}

It is a well known fact that the electronic properties of graphene depend strongly upon the deformation field applied to it, due, in part, to its high elastic response (about 23\% of the lattice parameter\cite{Lee08}). In fact, very interesting phenomena arise from applying different kinds of deformation fields. Among these phenomena we have band gap openings at the Fermi level\cite{Pereira09,PereiraLetter09}, shifts of the Dirac cones from their original positions\cite{Pereira09,Maurice}, localized energy edge modes\cite{Nosotros214,jose15}, fractal-like energy spectrum\cite{Nosotros14,Nosotros214,roman15}, merging of inequivalent Dirac cones\cite{Montambaux2008,Montambaux,Nosotros214,Feilhauer15}, tunable dichroism\cite{tunoliva15}, anisotropic AC conductivity\cite{Olivaany14}, new and interesting transport properties\cite{Babajanov2014,Mishra2015,Agarwala16,Carrillo16}, etc. All these have opened an avenue for the emergent field of straintronics\cite{Pereira09,Guinea12,Ni14,ChenSi16,Amorim16,Salary16}, which aim is to taylor the electronic properties of graphene via mechanical deformations. 

On the other hand, although graphene is a semimetal, it possesses non-trivial topological properties\cite{Volovik2011}. For instance, the zero-energy edge states observed in graphene are flat bands that join two inequivalent Dirac cones\cite{Montambaux2008}. Flat bands have its origin in the energy spectrum, which can host lines or points where bands touch each other at zero energy, as was first pointed out by Volovik\cite{volovikbook03,Volovik2011,Volovik2013}. This results from the Dirac equation topological properties. In fact, two inequivalent Dirac cones in graphene have opposite Berry phase. Since the states at the Dirac cone cannot be transformed into topologically trivial states (with Berry phase equal to zero), a flat band joining Dirac cones with opposite Berry phase emerges for a finite system \cite{Volovik2011}. The three dimensional (3D) version of Dirac semimetals (usually called Weyl semimetals) also gives rise to flat bands, known as Fermi arcs, joining Weyl points (points at zero energy where the bands cross each other) with opposite topological charge. These flat bands, as the ones that emerge in Dirac semimetals, are very stable, since both of them are protected by the bulk-edge correspondence\cite{Volovik2011}. This is a consequence of the fact that in the neighborhood of Weyl nodes, the effective Hamiltonian of the system can be described by a Weyl equation. Therefore, wave functions describe Weyl fermions with opposite chirality \cite{RaoWeyl16}, which means that the only way to open a gap is by the annihilation of two Weyl nodes with opposite chirality. Interestingly enough, recent experiments have shown Fermi arcs in real condensed matter systems \cite{Xu615,RaoWeyl16}.

The importance of flat bands stems from their potential to be used in technological applications as topological quantum computing\cite{kitaev01}. This is possible since Dirac and Weyl nodes always come in pairs and might have a Majorana-like nature\cite{toprev11,majsarma15,majgraphene15,kraus13}, which gives them robustness to weak perturbations and decoherence\cite{kitaev01}. 

Hence many theoretical condensed matter systems that exhibit topological edge modes have been proposed, among them, the most promising ones seem to be periodically driven systems, studied under the Floquet approach\cite{majss10,majpss13,Pikulin15,kitaev01,maj11,optlatt12,deng15,yang15,demlerqw11,thakurathi13,pedrocchi11, petrova14, clement14}. Actually, these system are able to host not only zero energy flat bands but also $\pm\pi$-energy flat bands\cite{demlerqw11,klinovaja16}. This results from the periodicity of the so called quasienergy spectrum, which arises in the frame of Floquet theory. Motivated by that, in this article, we study the case of time periodically uniaxial strained zigzag graphene nanoribbons (ZGNs) within the tight binding approach using the Floquet formalism, and, for the sake of simplicity, in the small strain's amplitude limit. We have found that the case system supports two kinds of zero-quasienergy flat bands and just one kind of $\pm\pi$ quasienergy flat bands. For the zero-quasienergy flat bands, we found that one is the well known zero edge state observed in pristine ZGNs, which is well understood in terms of flat bands joining two inequivalent Dirac cones with opposite chirality\cite{Volovik2011} or in terms of the Zak phase\cite{ZakPhase11}. The others arise as a consequence of the driving and can be understood as flat bands joining touching band points with opposite Berry phase. 

\begin{figure*}
\includegraphics[scale=0.65]{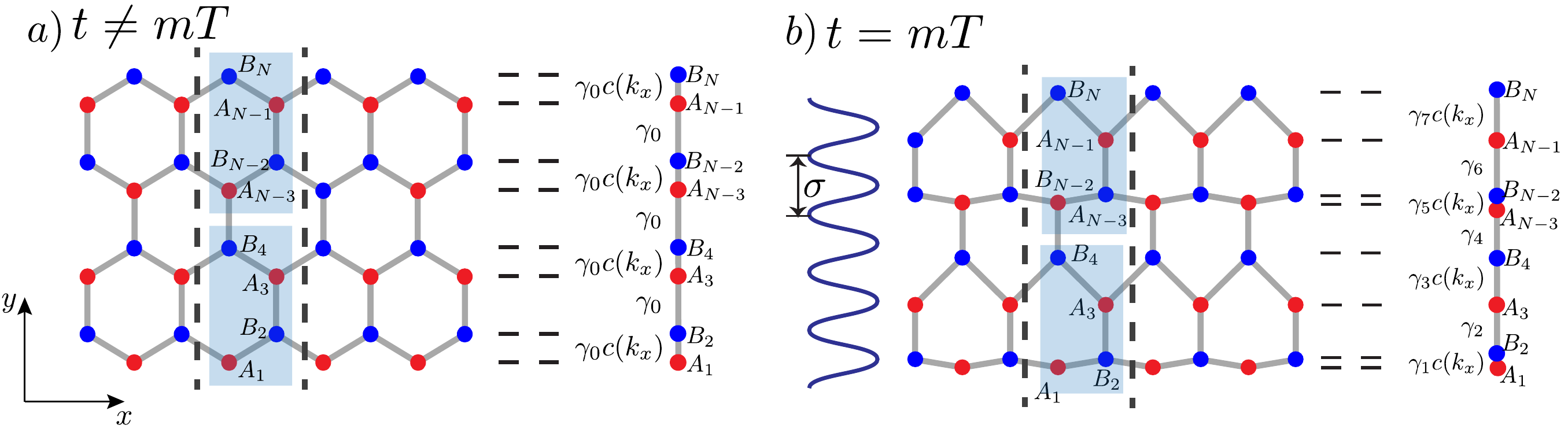}
\caption{(Color online). Layout of the periodically driven strained zigzag graphene nanoribbon. Basically, the strain field is turned off (see panel $a$) whenever that $t\neq mT$, where $T$ is the driving period and $m$ is an integer number. The strain field is turned on for $t=mT$, as shown in panel $b$. Since the strain field depends only upon the $y$-direction, the zigzag graphene nanoribbon can be mapped onto an effective one dimensional system, which is represented by linear chains in the figure. The dots indicate the position of the atoms on each graphene/linear chain row.}
\label{ZGN}
\end{figure*}

The layout of this paper is the following. First we present in Section \ref{model} the model, then in Section \ref{numerical} we present the quasienergy spectrum obtained from numerical results. 
Section \ref{analytical} is devoted to explain such results using an analytical approach based on an effective Hamiltonian. Section \ref{TBPs} contains an analysis of the analytical found spectrum and
the topological phase diagram. In Section \ref{topnat} we prove the non-trivial topological properties of the modes, while Section \ref{experiment} is devoted to an study of the experimental feasibility of our model. Finally, in Section \ref{conclusion} the conclusions are given. 

\section{Periodically driven strain graphene}
\label{model}

We start by considering a pristine zigzag graphene nanoribbon (ZGN) as the one displayed in Fig. \ref{ZGN} a). Then, we apply an uniaxial strain field along the $y$-direction $u(y)$ given by  
\begin{equation}
u(y)=\frac{2\lambda}{9}\cos{\left[\frac{8\pi}{3}\sigma(y-1/2)+\phi\right]}
\end{equation}
which is similar to the pattern of strain that emerges when graphene is growth on top of a different lattice substrate \cite{Nosotros14}. It is important to say that the strain field is tailored by three parameters, namely, the amplitude ($\lambda$), the frequency ($\sigma$) and, finally, the phase ($\phi$). Within the tight binding approach and considering the small strain's amplitude limit the electronic properties of an uniaxial strained ZGN are well described by the following effective one-dimensional (1D) Hamiltonian\cite{Nosotros14}
\begin{equation}
H(k_x)=\sum^{N-1}_{j=1} \left[\gamma_{2j}\,a_{2j+1}^{\dag} b_{2j}+c(k_x)\, \gamma_{2j-1} a_{2j-1}^{\dag} b_{2j}\right]+\mathrm{h.c.},
\label{1DsHam}
\end{equation}
where $c(k_x)=2\cos{\left(\sqrt{3}k_x/2\right)}$, $a$ is the interatomic distance between carbon atoms, $k_x$ is the crystal momentum in $x$-direction, $a_j$ ($b_j$) annihilates an electron at the $j$-th site in the sub lattice A (B) along the $y$-direction, and $N$ is the number of atoms within the unit cell (see Fig. \ref{ZGN}). Finally, the hopping parameters are given by
\begin{equation}
\gamma_j=\gamma_0+\lambda\,\gamma_0\xi(j+1)\sin{\left[\pi\sigma\xi(j)\right]}\sin{\left(2\pi\sigma j+\phi\right)},
\label{hop}
\end{equation}
where $\xi(j)=1+(-1)^{j}/3$ and $\gamma_0=2.3$ eV is the hopping parameter for unstrained graphene. Frequently, we will use $a$ (the interatomic distance between carbon atoms) as the unit of distance and $\gamma_0$ as the unit of energy, although, when necessary we will explicitly write them. Having said that, let us introduce the time dependence to the model. That will be done by considering the following driving layout 
\begin{equation}
\gamma_j(t)= \left\{\begin{array}{lll}
             \gamma_0 & if & t<\text{mod}(t,T)<t_1\\
             \gamma_j& if & t_1<\text{mod}(t,T)<T
            \end{array}\right.
\end{equation}
where $T$ is the period of the driving and $t_1$ is in the range $0<t_1<T$. This leads to the following time-dependent Hamiltonian 
\begin{equation}
\begin{split}
H(k_x,t)=\sum^{N-1}_{j=1} &\left[\gamma_{2j}(t)\,a_{2j+1}^{\dag} b_{2j}+c(k_x)\, \gamma_{2j-1}(t) a_{2j-1}^{\dag} b_{2j}\right]\\
&+\mathrm{h.c.}
\end{split}
\label{HZtime}
\end{equation}
The previous Hamiltonian describes a system for which the strain field is turned on during the interval $(t_1,T)$ and it is turned off whenever $t$ is on the range $(0,t_1)$. We will consider the case of short pulses, this is, $t_1\rightarrow T$. As long as the product of the kicking amplitude (here represented by the parameter $\lambda$, the strain's amplitude) and the duration of the pulse $T-t_1$ is kept constant, the kicking can be approximated by a Dirac delta function if the $t_1\rightarrow T$ limit is considered. This kind of kicking layout can be hard to be reached in experimental conditions, therefore, we discuss the experimental feasibility of our model in a special section (see section \ref{experiment}), therein, we also study a more realistic kind of driving: harmonic driving. However it is worth mentioning that many theoretical papers consider a quite similar kind of kicking\cite{Abal02,Creff06,thakurathi13,wangH13,Ho14,Bandy15,KHM16}. 

From here, we will study the $t_1\rightarrow T$ limit, then the driving protocol can be written as 
\begin{equation}
\begin{split}
&\gamma_j(t)=\gamma_0+\\
&\sum_m\delta(t/T-m)\gamma_0\lambda \xi(j+1) \sin\left[\pi \sigma \xi(j)\right] \sin(2 \pi \sigma j+\phi),
\label{gamdelta}
\end{split}
\end{equation}
where $m$ is an integer number and $T$ is the period of the driving. An schematic layout of the driving is shown in Fig. \ref{ZGN}. Therein, it can be seen that the strain field is turned on for $t=mT$ whereas is turned off for different times (this is, for $t\neq mT$).   

The advantage of considering kicking systems relies in the fact that the time evolution operator defined as 
\begin{equation}
U(T)\ket{\psi_{k}(t)}=\ket{\psi_k(t+T)},
\end{equation}
where $\ket{\psi_k(t)}$ is the wave function of the system for a given $k$, can be written in a very simple manner
\begin{equation}
\begin{split}
U(\tau)&=\mathcal{T}\exp{\left[-i\int_{0}^TH(k_x,t)\,dt/\hbar\right]}\\
 &=\exp{\left[-i\tau H_1\right]}\exp{\left[-i\tau H_0\right]},
\end{split}
\label{uop}
\end{equation}
where $\mathcal{T}$ denotes the time ordering operator, $\tau\equiv T/\hbar$, and
\begin{equation}
\begin{split}
H_0(k_x)&=\gamma_{0}\sum^{N-1}_{j=1} \left[a_{2j+1}^{\dag} b_{2j}+c(k_x)\,a_{2j-1}^{\dag} b_{2j}\right]+\mathrm{h.c.}\\
H_1(k_x)&=\sum^{N-1}_{j=1} \left[\delta\gamma_{2j} a_{2j+1}^{\dag} b_{2j}+c(k_x)\, \delta\gamma_{2j-1} a_{2j-1}^{\dag} b_{2j}\right]\\
 &+\mathrm{h.c.}
\end{split}
\end{equation}
with $\delta\gamma_j=\gamma_j-\gamma_0$. In general, Hamiltonians $H_1$ and $H_0$ do not commute, therefore, it is common to study the properties of the system through an effective Hamiltonian given by $U(\tau)=\exp{(-i\tau H_{\mathrm{eff}})}$, which has eigenvalues $\exp{(-i\tau\omega)}$, where $\tau\omega$ is called the quasienergy of the system. Note that the product $\tau\omega$ is defined up to integer multiples of $2\pi$ due to the periodicity of the Floquet space. Our periodically driven model Eq. (\ref{HZtime}) is very rich, since it has four parameters, three owing to the strain field ($\lambda$, $\sigma$, and $\phi$) and one to the driving ($\tau$). 

Even though one can study the system for different values of $\sigma$ and $\phi$ we will focus on the case $\sigma=1/2$ and $\phi=4\pi\sigma/3$, because this case has very interesting features and makes possible to perform analytical calculations. For these values of $\sigma$ and $\phi$, the hopping parameter takes the following form
\begin{equation}
\begin{split}
\gamma_{2j-1}-\gamma_0&=-\lambda\\
 \gamma_{2j}-\gamma_0&=\lambda/2.
\end{split}
\label{gammas}
\end{equation}
This means that the Hamiltonian $H_1$ is on a critical line that separates two distinct topological phases via the parameter $\lambda$ in the time-independent case. In such a case, for $\lambda<\lambda_{C}=0.4$, the system is on a non-trivial topological semimetal phase ({\it i.e.} the system is gapless, there are Dirac cones) and it is able to host edge modes \cite{Volovik2011}. For $\lambda>\lambda_{C}$ the system is on a normal Zak insulator phase (there are no Dirac cones and the system is gapped, however there still being zero energy edge states \cite{Montambaux2008,Montambaux}). It is interesting to see what happens at the critical value $\lambda_{C}$. At that point two inequivalent Dirac cones have merged and the dispersion relation has an anomaly, in the sense that it is quadratic in one direction, whereas in the other direction remains linear \cite{Nosotros214}. However, we have used the approximation of small strain's amplitude, so we are interested on $\lambda\ll\lambda_{C}$. The main reason for consider this is that provides a great simplification on theoretical calculations, moreover, it is much simpler to obtain small strain's amplitude in experimental setups. 

Once that the model has been described, the next step is to analyze the quasienergy spectrum as a function of $\tau$ (the driving period) keeping  $\sigma$, $\phi$, and $\lambda$ constant. The results of the numerical analysis, obtained by the numerical diagonalization of Eq. (\ref{uop}), are discussed  in the next section. 

\section{Quasienergy spectrum: Numerical results}
\label{numerical}
\begin{figure}
\includegraphics[scale=0.235]{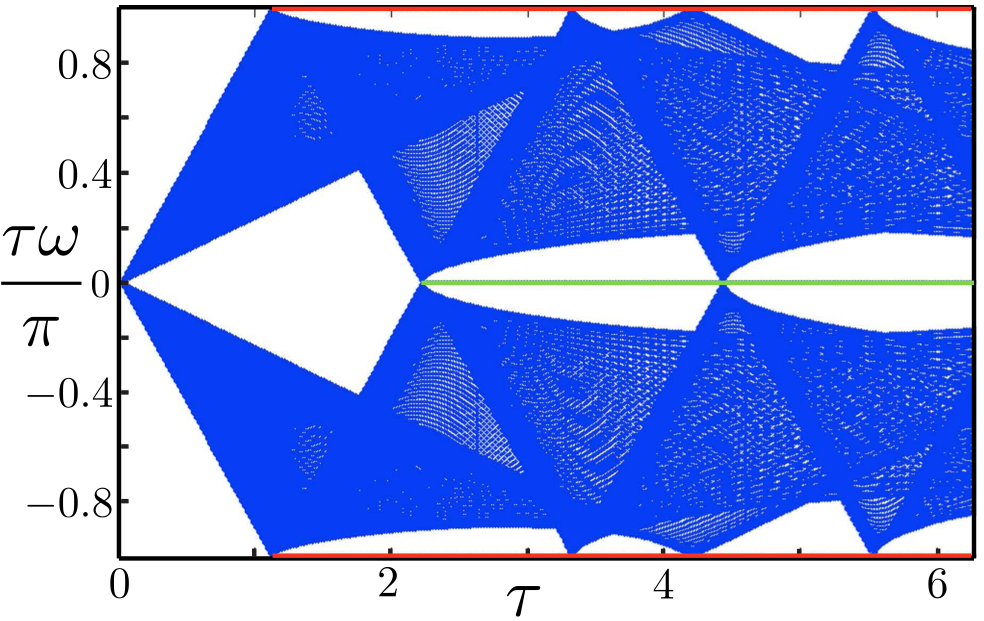}
\caption{(Color online). Quasienergy spectrum numerically obtained from the eigenvalues of the matrix representation of Eq. (\ref{uop}). For making the plot, we have used $k_x=0$, $\lambda=0.1$, $\sigma=1/2$, and $\phi=4\pi\sigma/3$, $N=240$ and fixed boundary conditions. Note that at certain values of $\tau$ the bands touch each other at $\tau\omega(0,k_y)=0,\,\pm\pi$. At such points flat bands emerge, indicated in the figure by red solid lines for $\tau\omega=\pm\pi$ and by green solid lines for $\tau\omega=0$. }
\label{phzig}
\end{figure}

We begin the study of the physic properties of the system by constructing the matrix representation of $U(\tau)$, Eq. (\ref{uop}), then we obtain its eigenvalues by numerical diagonalization. In all cases presented here we studied $\omega$ as a function of $k_x$ and $\tau$, using $\sigma=1/2$ and $\phi=4\pi\sigma/3$ for a system of $N=240$ sites per unit cell, and imposing fixed boundary conditions. The resulting quasienergy spectrum is shown in Fig. \ref{phzig} for a cut at $k_x=0$ using $\lambda=0.1$. For small $\tau$, the spectrum has a central gap that grows linearly with $\tau$. As can be seen in such figure, the outer band edges also grow linearly with $\tau$. Then, when $\tau$ reaches a critical value, denoted by $\tau_{c}$, the outer edge bands touch the limit of the first Brillouin zone of the Floquet space. At that point, flat bands emerge at $\pm\pi$ quasienergies, these bands are labeled by red solid lines in Fig. \ref{phzig}. If we continue increasing $\tau$, we will reach the point $\tau=2\tau_{c}$, at which the outer edge bands will touch each other again and a new flat band appears at zero quasienergy (denoted by green solid lines, see Fig. \ref{phzig}). The flat nature of these bands and the fact that they are separated by a finite gap from the other bands suggest that they are due to surface effects. Moreover, since these states emerge at crossing band points, they have a similar origin as the edge states that appear in the Shockley model\cite{Shockley39,davison92,Pershoguba12,deng15}, which always come in pairs and can have an exotic Majorana-like nature. Actually, these kind of edge states have been predicted to appear in a 1D s-wave superconductor wire\cite{maj11}. However, our system is two dimensional (2D), therefore we expect that edge modes that appear in Fig. \ref{phzig} give rise to flat bands in the band structure, each of these flat bands made out of Majorana-like modes.

\begin{figure}
\includegraphics[scale=0.315]{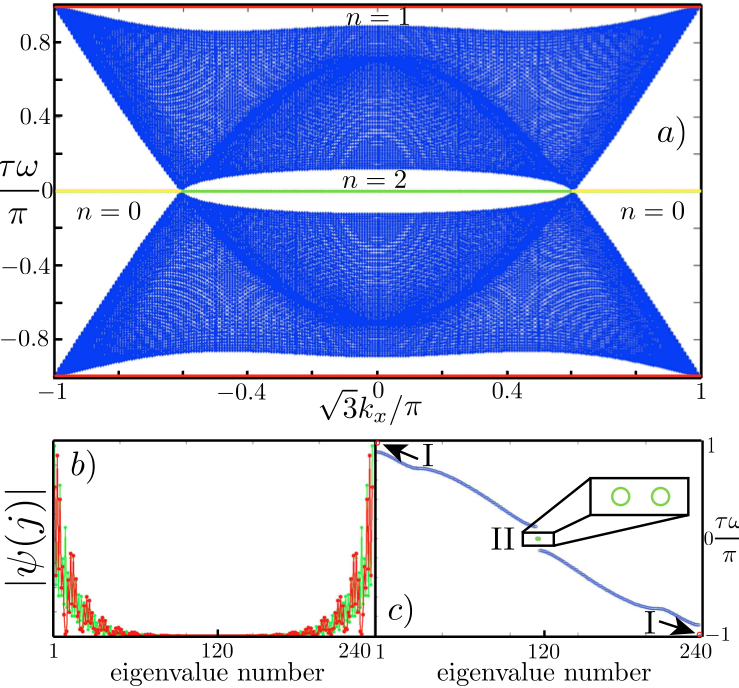}
\caption{(Color online). Upper panel. Quasienergy band structure as a function of $k_x$ for $\lambda=0.1$, $\sigma=1/2$, $\phi=4\pi\sigma/3$, and $\tau=3$. We have flat bands at zero and $\pm\pi$ quasienergies. Note that for $\tau\omega=0$ there are two types of flat bands, ones having a time-independent origin (yellow solid lines) and the others having a time-dependent origin (green solid lines), this is explained in the main text. The index $n$ indicates the corresponding region in the topological phase diagram and the types of edge states. For $n$ odd we have $\tau\omega=\pm\pi$ states (red color), while $n$ even indicates zero-quasienergy edge states (green color). The case $n = 0$ stands for time-independent edge states at $\tau\omega=0$ (yellow color). In panel b), two wave functions amplitude for $\tau\omega=0$ and $\tau\omega=\pi$ using $k_x=0$ are shown. The amplitudes follow the same color code as in panel a). Panel c), the quasienergy value is presented as a function of the quasienergy eigenvalue number for $k_x=0$.}
\label{bszt}
\end{figure}

\begin{figure}
\includegraphics[scale=0.32]{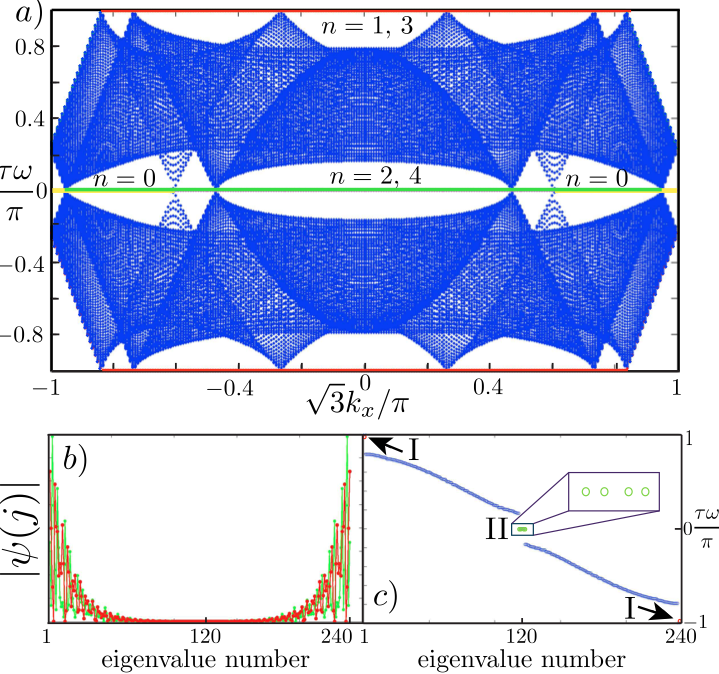}
\caption{(Color online). Upper panel. Quasienergy band structure, made under the same conditions of Fig. \ref{bszt} but using $\tau=5.28$. The label $n$ indicates the corresponding region in the topological phase diagram and the types of edge states. For $n$ odd we have $\tau\omega=\pm\pi$ states (red color), while $n$ even indicates zero-quasienergy edge states (green color). The case $n = 0$ stands for time- independent edge states at $\tau\omega=0$ (yellow color). Panel b), we show the wave functions amplitudes for edge states at $k_x=0$ using the same color code as in panel a). In c), we present the quasienergy value as a function of the number of quasienergy eigenvalue.}
\label{bszf}
\end{figure}

To confirm the previous conjecture, we plotted the quasienergy spectrum as a function of $k_x$ for $\tau=3$ (see Fig. \ref{bszt}) and $\tau=5.28$ (see Fig. \ref{bszf}) under the same conditions of Fig. \ref{phzig}. In panels b) of Figs. \ref{bszt} and \ref{bszf} we show the amplitude of the wave functions with flat dispersion for $k_x=0$. Note that these states are localized near the edges of the unit cell and that they come in pairs. Additionally there is a finite gap (although not a full gap) that separates such states from the rest bands, which suggests that they have non-trivial topological properties and that they posses a Majorana-like nature. Furthermore, we can see three kinds of edge states, one at $\pm\pi$ quasienergy (indicated by I in solid red lines) and the others as zero quasienergy (indicated by II in yellow and green solid lines). The yellow flat bands, as we will discuss below, are the well known zero edge modes that emerge in a finite pristine ZGN due to edge effects and have nothing to do with the driving, whereas the other ones (the green and red ones) are a consequence of the driving. It is important to mention that flat bands are very robust under the driving. Note that flat bands always emerge from touching band points either at $\pm\pi$ or zero quasienergy, which suggests that the origin of them is quite similar to that of Fermi arcs, which join two different Weyl points ({\it i.e.} points on the momentum space at where energy vanishes) with opposite chirality \cite{Burkov11}. To confirm or refuse that conjecture a more detailed analysis is required. The next section is devoted to that aim.

%

\section{Analytical study of the quasienergy spectrum}
\label{analytical}

\begin{figure}
\includegraphics[scale=0.345]{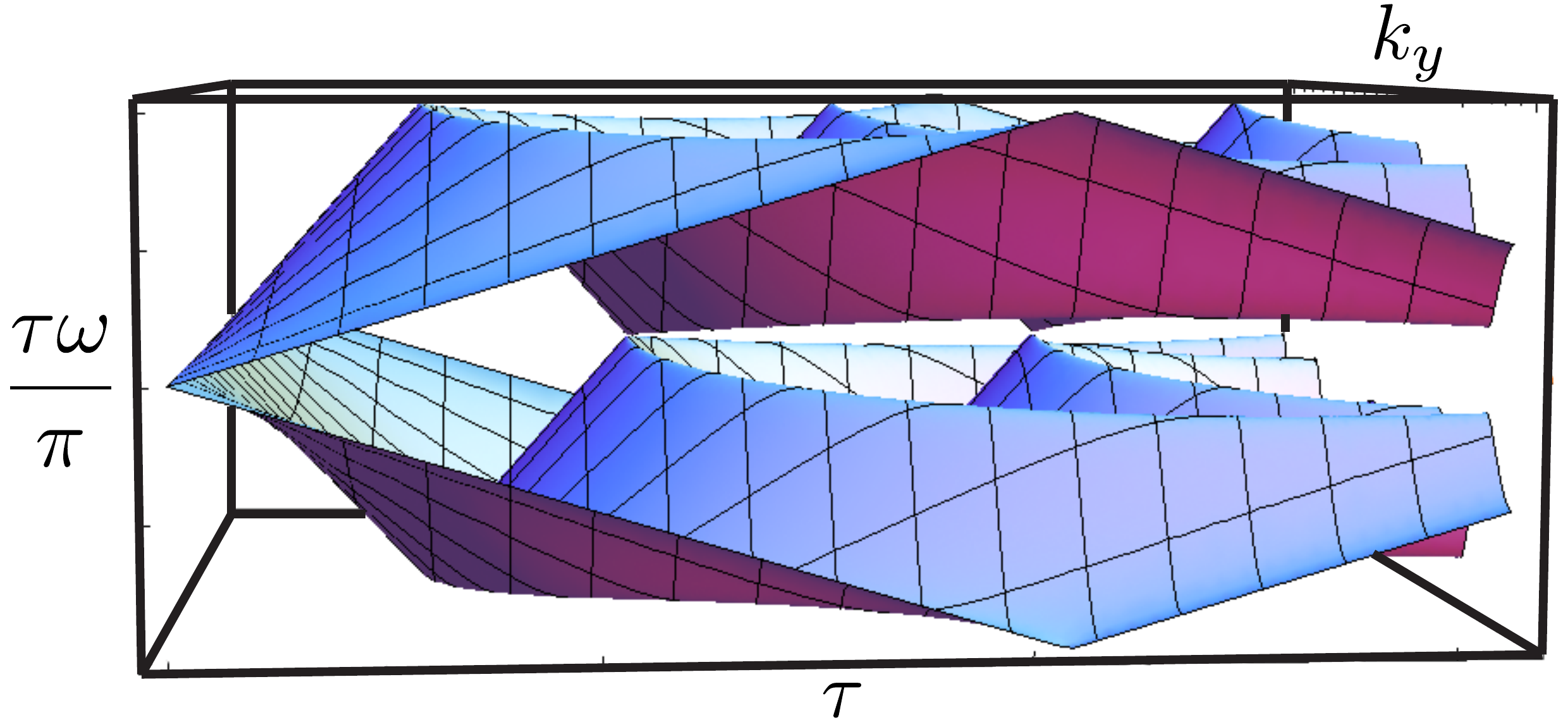}
\caption{(Color online). Analytical quasienergy spectrum obtained from Eq. (\ref{omeganew}). In the vertical axis we plot $\tau \omega/\pi$ as a function of $k_y$ and $\tau$ for $k_x=0$, $\lambda=0.1$, $\sigma=1/2$, and $\phi=2\pi/3$. Note that this figure reproduces the quasienergy spectrum obtained numerically by a diagonalization of the Hamiltonian, as shown in Fig. \ref{phzig}. However, the flat bands that appear in Fig \ref{phzig} are missing here since this is a surface effect.}
\label{phda}
\end{figure}

Once the numerical results have been stablished, we will proceed to explain them analytically. This will be done by studying the quasienergy spectrum for $\sigma=1/2$ and $\phi=4\pi\sigma/3$, imposing cyclic boundary conditions in the $y$-direction. This is possible because for $\sigma=1/2$ the hopping parameters just take two different values (see Eq. (\ref{gammas})), therefore the system becomes periodic in the $y$-direction and $k_y$ is a good quantum number. We proceed as usual, {\it i.e.}, first, we define the following Fourier transform for the annihilation operators 
\begin{equation}
\begin{split}
a_j&=\frac{1}{\sqrt{N/2}}\sum_{k_y}e^{-i3k_y j/2}a_{k_y}\\
b_j&=\frac{1}{\sqrt{N/2}}\sum_{k_y}e^{-i3k_y j/2}b_{k_y}.
\label{fourier}
\end{split}
\end{equation}
and apply them into Hamiltonians $H_1$ and $H_0$, Eq. (\ref{hams}). It is straightforward to show that the bulk Hamiltonians are given by 
\begin{equation}
\begin{split}
H_0(k_x,k_y)&=h_0(k_x,k_y)\,\mathbf{\hat{h}_0}\cdot\mathbf{\sigma}\\
H_1(k_x,k_y)&=h_1(k_x,k_y)\,\mathbf{\hat{h}_1}\cdot\mathbf{\sigma}
\end{split}
\label{hams}
\end{equation}
where $\sigma_i$ ($i=x\,,y\,,z$) is a $2\times2$ Pauli matrix defined in the basis where $\sigma_z$ is diagonal. The components of $\mathbf{h_0}$ and $\mathbf{h_1}$ are
\begin{equation}
\begin{split}
h_0^{(x)}(k_x,k_y)&=2\cos{\left(\sqrt{3}k_x/2\right)}+\cos{\left(3k_y/2\right)}\\
h_0^{(y)}(k_x,k_y)&=\sin{\left(3k_y/2\right)},\\
h_1^{(x)}(k_x,k_y)&=-2\lambda\cos{\left(\sqrt{3}k_x/2\right)}+\frac{\lambda}{2}\cos{\left(3k_y/2\right)}\\
h_1^{(y)}(k_x,k_y)&=\frac{\lambda}{2}\sin{\left(3k_y/2\right)}.
\end{split}
\end{equation}
From this we define the norms $h_0=\left |\mathbf{h_0}\right|$ and $h_1=\left |\mathbf{h_1}\right|$. Therefore, the time evolution operator, Eq. (\ref{uop}), is given by
\begin{equation}
\mathcal{U}(k_x,k_y,\tau)=\exp{\left[-i\tau H_1(k_x,k_y)\right]}\exp{\left[-i\tau H_0(k_x,k_y)\right]}
\label{eqUeff}
\end{equation}
where $U(\tau)=\sum_{k_y}\mathcal{U}(\tau,k_x,k_y)\otimes \ket{k_y}\bra{k_y}$. The Hamiltonians $H_1(k_x,k_y)$ and $H_0(k_x,k_y)$ do not commute since (see Appendix A)
\begin{equation}
\left[H_1,H_0\right]=-6i\lambda\sin{\left(3k_y/2\right)}\cos{\left(\sqrt{3}k_x/2\right)}\sigma_z.
\label{comm}
\end{equation}
Yet, it is still being possible to write,
\begin{equation}
\mathcal{U}(k_x,k_y,\tau)=\exp{\left[-i\tau H_{\mathrm{eff}}(k_x,k_y)\right]}.
\label{uopeff}
\end{equation}
Using the results obtained in Appendix A, the effective Hamiltonian $H_{\mathrm{eff}}(k_x,k_y)$ can be written as 
\begin{equation}
H_{\mathrm{eff}}(k_x,k_y)=\omega(k_x,k_y)\,\mathbf{\hat{h}_{\mathrm{eff}}}\cdot\mathbf{\sigma},
\label{HZeffT}
\end{equation}
where $\mathbf{\hat{h}_{\mathrm{eff}}}$ is a unit vector (whose explicit form is also given in Appendix A). The quasienergies of the system, $\pm\tau\omega(k_x,k_y)$, are given by (see Appendix A)
\begin{equation}
\begin{split}
&\cos{\left[\tau\omega(k_x,k_y)\right]}=\cos{(\tau h_1)}\cos{(\tau h_0)}\\
&-\mathbf{\hat{h}_{0}}\cdot\mathbf{\hat{h}_{1}}\sin{(\tau h_1)}\sin{(\tau h_0)}
\end{split}
\label{omeganew}
\end{equation}
with
\begin{equation}
\begin{split}
&\mathbf{\hat{h}_{0}}\cdot\mathbf{\hat{h}_{1}}=\frac{\lambda}{h_1h_0}\times\\
&\left[-4\cos^2{\left(\sqrt{3}k_x/2\right)}-\cos{\left(\sqrt{3}k_x/2\right)}\cos{\left(\frac{3k_y}{2}\right)}+\frac{1}{2}\right].
\end{split}
\end{equation}
Through Eq. (\ref{omeganew}) we are able to exactly reproduce the quasienergy bands obtained by numerical calculations. For example, in Fig. \ref{phda} we plot $\omega(0,k_y)$ obtained from Eq. (\ref{omeganew}), showing an excellent agreement with its  numerical counterpart displayed in Fig. \ref{phzig}. Observe that cyclic boundary conditions were used for obtaining Fig. \ref{phda}, and thus the edge states seen in Fig. \ref{phzig}  do not appear. 
%

\section{Touching band points}
\label{TBPs}

Since flat bands emerge from touching band points at $\tau\omega=n\pi$ ($n$ an integer number), knowing its exact location is crucial.  This is the subject of the present section.
We start by observing that touching band points are obtained by setting $\tau\omega=n\pi$ in Eq. (\ref{omeganew}), resulting   in the condition,

\begin{equation}
\begin{split}
&\pm1=\cos{(\tau h_1)}\cos{(\tau h_0)}\\
&-\mathbf{\hat{h}_{0}}\cdot\mathbf{\hat{h}_{1}}\sin{(\tau h_1)}\sin{(\tau h_0)}
\end{split}
\label{Maxomeganew}
\end{equation}
where it is understood that the previous condition holds only for touching bands points. We will denote such special $k$ points by using a star, {\it i.e.}, $(k_x^{*},k_y^{*})$.
A detailed analysis shows that Eq. (\ref{Maxomeganew}) is satisfied for two possible cases,

\begin{itemize}

\item[1.] The first one requires that $\mathbf{\hat{h}_{0}}\cdot\mathbf{\hat{h}_{1}}=\pm1$. This is equivalent to ask $\mathbf{\hat{h}_{0}}\times \mathbf{\hat{h}_{1}}=0$. Since 
$[H_0,H_1]=-3i\,h_0h_1\left(\mathbf{\hat{h}_{0}}\times \mathbf{\hat{h}_{1}}\right)\cdot\hat{e}_z \,\sigma_z$, the condition is equivalent to $[H_0,H_1]=0$.

\item[2.] The second case is $\mathbf{\hat{h}_{0}}\cdot\mathbf{\hat{h}_{1}}\neq\pm1$, which is equivalent to $[H_0,H_1]\neq 0$. However, in this case it is required  the extra condition $\cos{(\tau h_1)}\cos{(\tau h_0)}=\pm1$ .  

\end{itemize}

As we will see later on, the first case  $\mathbf{\hat{h}_{0}}\cdot\mathbf{\hat{h}_{1}}=\pm1$ gives rise to edge states, which are flat bands that join a kind of Weyl nodes with opposite Berry phase. They can emerge for small strain's amplitudes. Although the second case  $\mathbf{\hat{h}_{0}}\cdot\mathbf{\hat{h}_{1}}\neq\pm1$ also hosts edge states, such states are no longer flat bands, instead their quasienergy varies with $k_x$. Unfortunately, the last kind of edge states emerge for big strain amplitude, which make them hard to be observed. As a consequence, we will find the location of such second case points, but we will focus only on the topological modes resulting from the first kind of touching band points.

\subsection{Touching band points for $\mathbf{\hat{h}_{0}}\cdot\mathbf{\hat{h}_{1}}=\pm1$}

From Eq. (\ref{omeganew}) we find that $\mathbf{\hat{h}_{0}}\cdot\mathbf{\hat{h}_{1}}=\pm1$ only if $k_x^{*}=\pi/\sqrt{3}$ or $k_y^{*}=0,\,\pm 2\pi/3$. It can be proved that the solution for $k_x^{*}=\pi/\sqrt{3}$ is contained in the ones for $k_y^{*}=0,\,\pm 2\pi/3$. Thus, we only analyze the cases $k_y^{*}=0,\,\pm 2\pi/3$. By substituting $k_y^{*}$ into Eq. (\ref{omeganew}),
\begin{equation}
\tau\omega_{\pm}(k_x)=\tau(1+\lambda/2)\pm2\tau(1-\lambda)\cos{\left(\sqrt{3}k_x/2\right)},
\label{omegapm}
\end{equation}
where the `$+$' sign stems for $k_y=0$ and the `$-$' sign for $k_y=\pm2\pi/3$. Now we require the condition  $\tau\omega_{+}(k_x)=n\pi$ (with $n$ an integer number) in Eq. (\ref{omegapm}) at a special $k_x=k_x^{*}$. This gives two possible values for $k_x^{*}$
\begin{figure}
\includegraphics[scale=0.38]{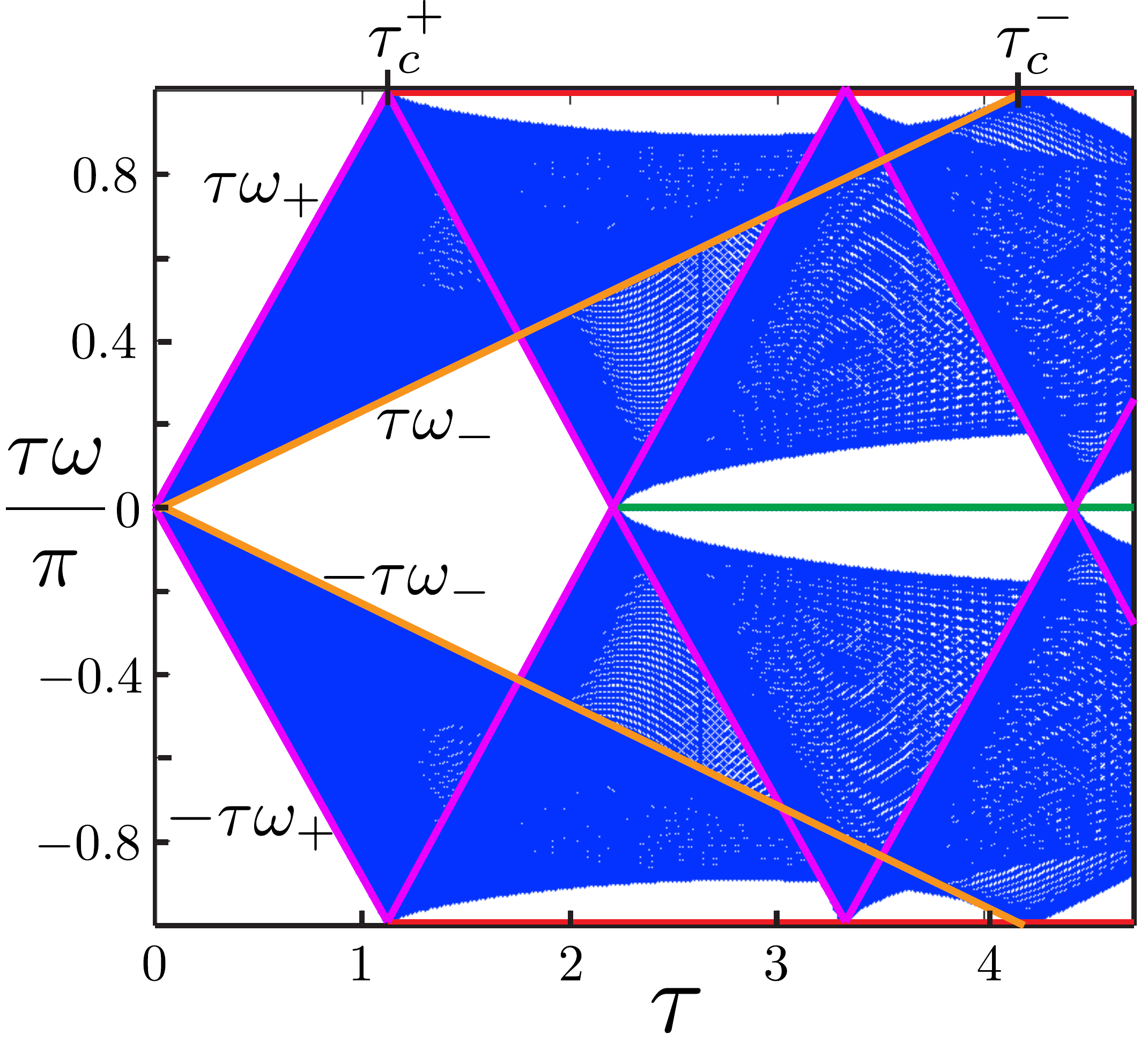}
\caption{(Color online). Band edges of the quasienergy spectrum as a function of $\tau$, calculated using the same conditions as in Fig. \ref{phzig}. The upper limits are indicated by pink solid lines and labeled by $\pm\tau\omega_{+}$, whereas the lower limits are shown by orange solid lines and labeled by $\pm\tau\omega_{-}$. Both limits, $\pm\tau\omega_{+}$ and $\pm\tau\omega_{-}$, were found from Eq. (\ref{omegapm}). The limits touch each other at $\tau_c=n\tau_{c}^{+}$ or $\tau_c=n\tau_{c}^{-}$ as indicated. It is clear that edge states emerge when two different bands touch each other, therefore, these states have a Shockley like nature\cite{Shockley39,davison92,Pershoguba12,deng15}.}
\label{phdr}
\end{figure}
\begin{equation}
\begin{split}
k_x^{*(+)}&=\pm\frac{2}{\sqrt{3}}\arccos{\left[\frac{ n\pi/\tau-(1+\lambda/2)}{2(1-\lambda)}\right]}\\
k_x^{*(-)}&=\pm\frac{2}{\sqrt{3}}\arccos{\left[\frac{ -n\pi/\tau+(1+\lambda/2)}{2(1-\lambda)}\right]}.
\end{split}
\label{kx}
\end{equation}
As before, $k_x^{*(+)}$ stems for $k_y^{*}=0$ and $k_x^{*(-)}$ for $k_y^{*}=\pm2\pi/3$. Note that  equation (\ref{kx}), for a given $n$, has two different solutions for $k_x^{*(+)}$ and four solutions for $k_x^{*(-)}$. It is noteworthy that since the cosine function is bounded, such solutions will exist and be real if and only if,
\begin{equation}
\left|\frac{n\pi/\tau-(\lambda+1/2)}{2(1-\lambda)}\right|\leq 1.
\label{phased}
\end{equation}
From the previous equation, we can obtain the minimum or critical value of $\tau$ for having touching band points at $\tau\omega=\pm n\pi$. Since we are looking for the minimum value of $\tau$ needed to have touching band points, it is enough to consider the equality in Eq. (\ref{phased}). If $\tau_c$ is the value at which the equality in Eq. (\ref{phased}) is held, we have that
\begin{equation}
\frac{n\pi/\tau_c\pm(\lambda+1/2)}{2(1-\lambda)}=\mp1.
\end{equation}
Two kinds of critical values of $\tau_c$ are obtained. Either
 $\tau_c=n\tau_{c}^{+}$ or $\tau_c=n\tau_{c}^{-}$, with
\begin{equation}
\tau_{c}^{+}=\frac{2\pi}{3(2-\lambda)}
\label{taumin}
\end{equation}
and
\begin{equation}
\tau_{c}^{-}=\frac{2\pi}{\left|5\lambda-2\right|}.
\label{tauplus}
\end{equation}
Now we explain why there are two critical values of $\tau$. Basically, $n\tau_{c}^{+}$  gives the touching band points that arise from 
the crossings between $\pm\tau\omega_{+}(k_x)$,  as indicated in Fig. \ref{phdr} for 
the quasienergy spectrum as a function of $\tau$ for $\lambda$ fixed and $k_x=0$. It is important to say that 
whenever $\tau$ reaches a critical value $n\tau_c^{+}$, a new pair of touching band points appear. Notice that this argument explains the
shape of the plot presented for the numerical results of Fig. \ref{phzig}. From Figs. \ref{phzig} and \ref{phdr}, is clear that edge states 
emerge when two different bands touch each other. These states have a Shockley like nature\cite{Shockley39,davison92,Pershoguba12,deng15}.


In a similar way, if $\tau$ is increased from zero, the quasienergies $\pm\tau\omega_{-}(k_x)$ will reach the edges of the Floquet space. This will happen at $\tau_{c}^{-}$, where $\tau_{c}^{-}>\tau_{c}^{+}$, see Fig. \ref{phdr}. As before, if $\tau$ increases up to $2\tau_c^{-}$, then $\tau\omega_{-}$ and $-\tau\omega_{-}$ will touch each other at zero quasienergy. New touching band points will appear each time that $\tau$ reaches $n\tau_{c}^{-}$.

\begin{figure}
\includegraphics[scale=0.256]{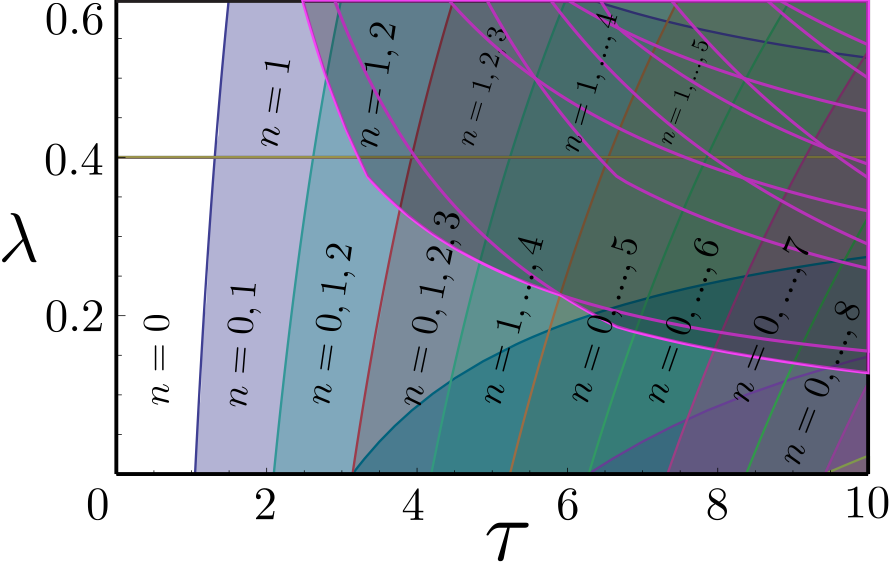}
\caption{(Color online.) Topological phase diagram, where the colors indicate regions of different maximal allowed $n$. Here the number of  topological non-trivial edge states increases with $n$. The phase boundaries are determined from Eq. (\ref{taumin}) and Eq. (\ref{tauplus}), using $\tau=n\tau_{c}^{+}$ and $\tau=n\tau_{c}^{-}$. The shaded region with magenta lines corresponds to non-flat band phases given by the condition $\mathbf{\hat{h}_{0}}\cdot\mathbf{\hat{h}_{1}}\neq1$. Phases with $\lambda<0.4$, as indicated by the horizontal line, are non-gapped at zero quasienergy for $\tau<2\tau_{c}^{+}$.}
\label{tdph}
\end{figure}

 Therefore, the number of pairs of touching band points will depend upon $\tau$ and $\lambda$. By plotting Eq. (\ref{phased}) for different values of $n$, the phase diagram of the system can be built.  In Fig. \ref{tdph}, such diagram is displayed. Therein, each color represents a phase of the system with the indicated allowed values of $n$. For instance, for $\lambda\leq 0.4$, the white color indicates just two pair of touching band points, since only one value of $n$ is allowed. On the other hand, for the violet color and $\lambda\leq0.4$, there are two touching band points pairs since $n=0,1$, or in other words, there are two allowed values for $n$. 

Up to now, we have found the location of touching band points at $\tau\omega(k_x^{*},k_y^{*})=\pm n\pi$, but a more detailed analysis is needed since two cases are of great interest.  Firstly, the case $n=0$, which give rise to touching band points at zero quasienergy at any value of $\tau$, suggesting that such points have a time-independent origin. Secondly, $n\neq0 $, {\it i.e.} touching band points at zero or $\pm\pi$ quasienergy. The emergence of such points depend upon the value of $\tau$ and $\lambda$ as can be seen in Fig. \ref{tdph}. 

%
\begin{figure*}
\includegraphics[scale=0.56]{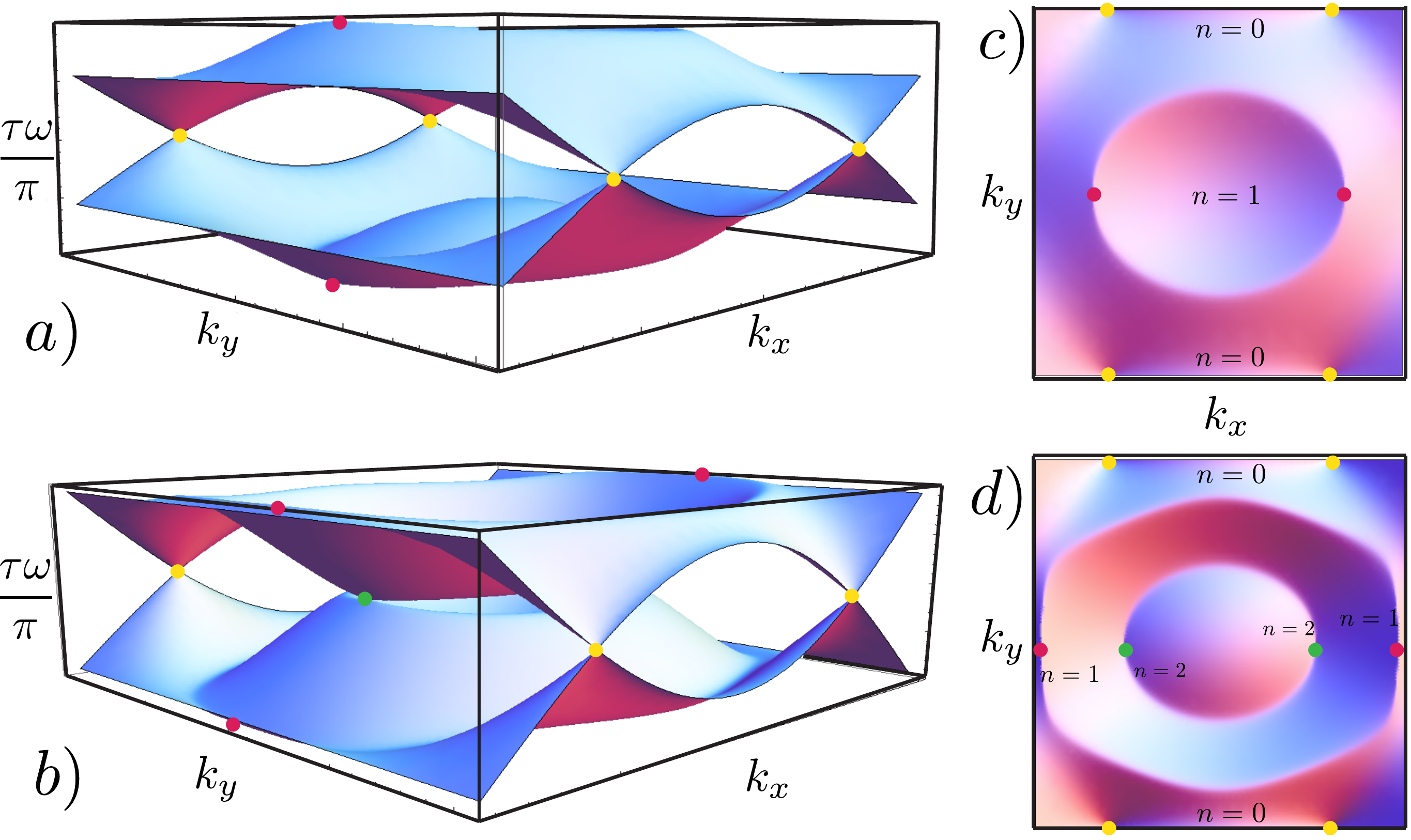}
\caption{(Color online.) Left panel. Band structure obtained using the analytical effective Hamiltonian quasienergies given by Eq. (\ref{omeganew}) for, a) $\tau=1.5\tau_c^{+}$, and b), $\tau=2.5\tau_c^{+}$ using $\lambda=0.1$. On the right, in panels c) and d) we show upper views of the same band structures. Therein, the touching band points are clearly seen. In panel c), corresponding to $\tau=1.5\tau_c^{+}$ there are two pairs of touching band point for $\tau\omega=0$ and another pair at $\tau\omega=\pm\pi$, which are denoted by yellow and red dots, respectively. As is proven in the main text, the yellow dots are Dirac cones vertices, which have a time-independent origin. On the other hand, red touching band points have a time-dependent origin. For $\tau=2.5\tau_c^{+}$ (see panel b)), the touching band points are at $\tau\omega=0$ (label $n=2$) and at $\tau\omega=\pm\pi$ (label $n=1$). The Dirac vertices remain the same as in panel a), corresponding to $n=0$.}
\label{vistas}
\end{figure*} 
\begin{figure*}
\includegraphics[scale=0.56]{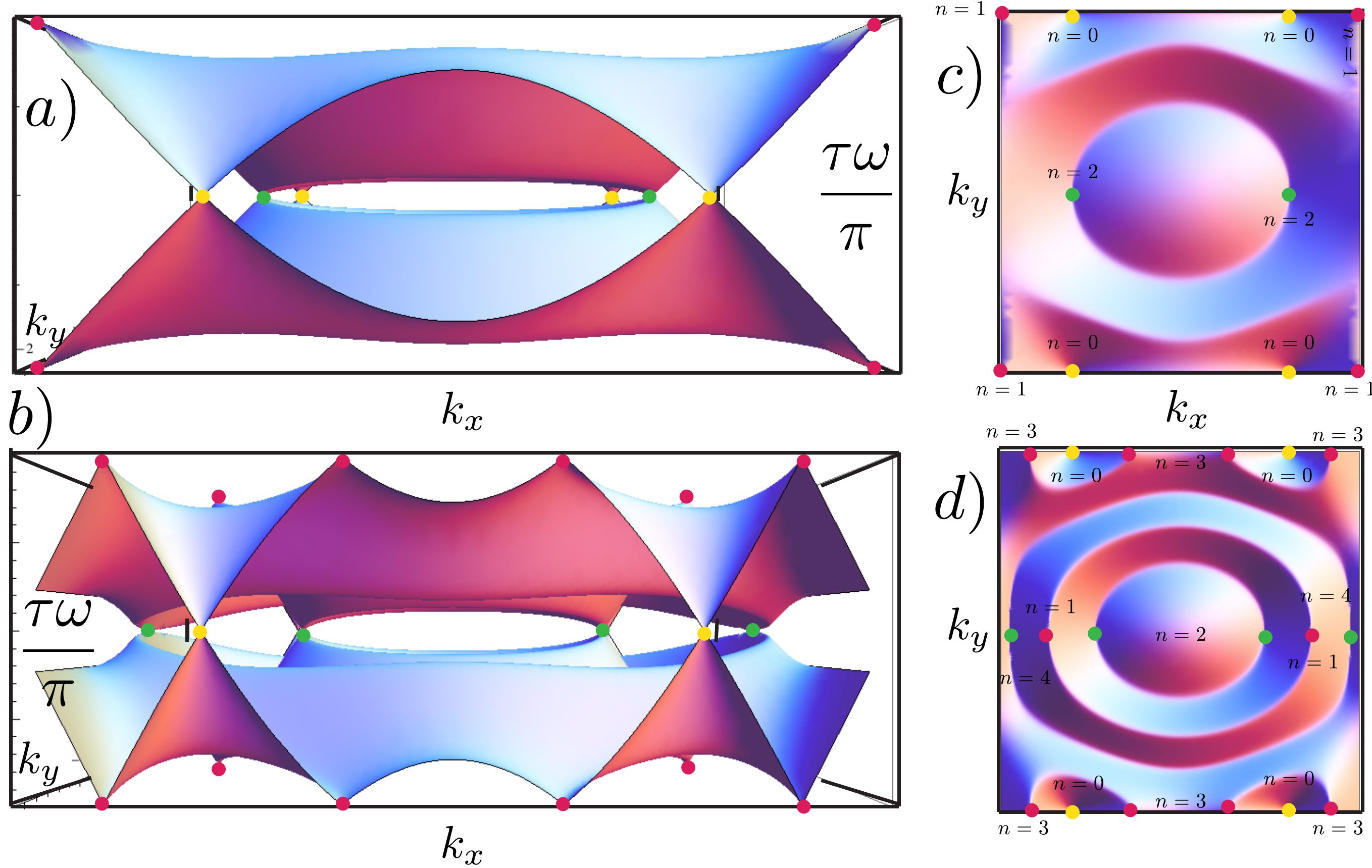}
\caption{(Color online.) Left panel. Band structure obtained using the analytical effective Hamiltonian quasienergies given by Eq. (\ref{omeganew})  for a) $\tau=3$ and b) $\tau=5.28$ using the same conditions as in Fig. \ref{vistas}, $\lambda=0.1$. On the right panel, upper views of the same band structure are shown. Note that in panel a), we have $\tau<\tau_{c}^{-}$, therefore there are two pairs of touching band points at $\pm\pi$ quasienergy. On the other hand, in panel b) we have $\tau>\tau_{c}^{-}$ and two new pairs of touching band points appear at quasienergy $\pm\pi$, see main text.  The parameters used for making this plot are the same than those used in Figs. \ref{bszt} and Figs. \ref{bszf}, in which a numerical diagonalization of the Hamiltonian was performed. This plot confirms that the numerical and analytical calculations are in excellent agreement.}
\label{vistados}
\end{figure*} 
%


First we will study {\it time-independent touching band points}. By setting $n=0$ in Eq. (\ref{kx}), we obtain
\begin{equation}
\begin{split}
k_x^{*(-)}&=\pm\frac{2}{\sqrt{3}}\arccos{\left[\frac{1+\lambda/2}{2(1-\lambda)}\right]}.
\end{split}
\label{kx0}
\end{equation}
Therefore, there are two touching band points pairs for $n=0$, one pair for each value of $k_y^{*}$, both located at $\pm k_x^{*}$. Moreover, from Eq. (\ref{kx0}), we found that these points are Dirac cones shifted from their original position due to the strain field. As we will see in the next section, this kind of touching band points will give rise to flat bands if the system is considered to be finite. For illustrating purposes, in Figs. \ref{vistas} and \ref{vistados} we present the band structure obtained using the analytical effective Hamiltonian quasienergies given by Eq. (\ref{omeganew}). Therein, the Dirac cones for $n=0$ are indicated by yellow points. 

It is important to say that Dirac cones undergo a phase transition as $\lambda$ is increased in the time-independent case. For $\lambda<\lambda_{C}=0.4$ there are two Dirac cones, indicated in Fig.  \ref{tdph} by a horizontal line at $\lambda_C$. When $\lambda$ reaches $\lambda_{C}$, the Dirac cones merge at a single point and, finally, for $\lambda>\lambda_{C}$ the energy spectrum becomes gapped. 

Here we are interested just in $\lambda\ll\lambda_{C}$, hence the gap opening is far away from this limit. Additionally, our system cannot become gapped since for $\tau\geq2\tau_{c}^{+}$, touching band points will emerge at zero quasienergy, avoiding the opening of a fully gap.


Second, we study  the {\it time-dependent touching band points} ($n \neq 0$). Two different types of touching band points emerge depending on the value of $n$. Since for touching band points we have that $\tau\omega(k_x{*},k_y^{*})=n\pi$, it follows that $\mathcal{U}(k_x^{*},k_x^{*},\tau)=(-1)^n$. For odd $n$, we have $\mathcal{U}(k_x^{*},k_x^{*})=-1$, this means that, due to the Floquet periodicity, touching band points at $\pm n\pi$-quasienergy ($n$ being an odd integer) are equivalent to touching band points at $\pm\pi$ quasienergy. Similarly, for even $n$ we have $\mathcal{U}(k_x^{*},k_x^{*})=1$, which implies that touching band points at $\pm n\pi$ quasienergy ($n$ being an even integer) are equivalent to touching band points at zero quasienergy. In Figs. \ref{vistas}, and \ref{vistados}, we labeled touching band points for odd $n$ by red dots, whereas touching band points for even $n$ are labeled by green points. The touching band points always come in pairs for a given value of $n$, as can be inferred from Eq. (\ref{kx}).  These different kinds of points, lead to different edge states as indicated in Figs. \ref{bszt} and \ref{bszf}. Therein, green flat bands result from joining a pair of touching band points for  even $n$. Red flat bands join pairs of odd $n$ touching band points.

\subsection{Touching band points for $\mathbf{\hat{h}_{0}}\cdot\mathbf{\hat{h}_{1}}\neq\pm1$}

Let us start by finding the location of these kind of touching band points. We first set $\cos{\left(\tau h_1\right)}=m_1 \pi$ and $\cos{\left(\tau h_0\right)}=n_1\pi$, where $m_1$ and $n_1$ are integer numbers. Then, after some algebraic operations, one gets
\begin{equation}
\begin{split}
k_y^{*}&=\frac{2}{3}\cos^{-1}{\left[\frac{\frac{\pi^2n_1^2}{6\tau^2}-\frac{m_1^2\pi^2}{6\tau^2\lambda^2}-\frac{1}{8}}{\sqrt{\frac{\pi^2}{3\tau^2}\left(\frac{m_1^2}{2\lambda^2}+\frac{n_1^2}{4}\right)-\frac{1}{8}}}\right]}\\
k_x^{*}&=\frac{2}{\sqrt{3}}\cos^{-1}{\left[\sqrt{\frac{\pi^2}{3\tau^2}\left(\frac{m_1^2}{2\lambda^2}+\frac{n_1^2}{4}\right)-\frac{1}{8}}\right]}
\end{split}
\end{equation}
In order to have real-valued $k_x^{*}$ and $k_y^{*}$, the following conditions must be fulfilled altogether
\begin{equation}
\begin{split}
&0\leq \frac{\pi^2}{3\tau^2}\left(\frac{m_1^2}{2\lambda^2}+\frac{n_1^2}{4}\right)-\frac{1}{8}\leq1 \\
&\left|\frac{\frac{\pi^2n_1^2}{6\tau^2}-\frac{m_1^2\pi^2}{6\tau^2\lambda^2}-\frac{1}{8}}{\sqrt{\frac{\pi^2}{3\tau^2}\left(\frac{m_1^2}{2\lambda^2}+\frac{n_1^2}{4}\right)-\frac{1}{8}}}\right|\leq1.
\end{split}
\label{cons}
\end{equation}
Therefore, the phase diagram shown in Fig. \ref{tdph} has to be modified, since the previous constrictions add new phases to the system. In the phase diagram shown in Fig. \ref{tdph}. These new phases appear in the shadowed area. The different phases are separated by the magenta curves. However, such values of strain are  difficult to achieve so in the present work we skip the analysis of their topological properties.  

\section{Topological nature of edge states}
\label{topnat}

The topological characterization of the flat bands for $\mathbf{\hat{h}_{0}}\cdot\mathbf{\hat{h}_{1}}=\pm1$ will be done in this section. To do that we will calculate the Berry phase around  the touching band points found before. The Berry phase is defined as
\begin{equation}
\gamma_{C}=\oint_{C}\mathbf{A}\cdot\,d\mathbf{k}
\label{Berry}
\end{equation}
where $\mathbf{A}=-i\bra{\psi_k}\mathbf{\nabla}_{k}\ket{\psi_k}$ is the so-called Berry connection (a gauge invariant quantity), and $\nabla_{k}=(\partial_{k_x},\partial_{k_y})$ is the gradient operator in the momentum space. We follow a four steps method to calculate such quantity. First, we note that exactly at the touching band points with $\mathbf{\hat{h}_{0}}\cdot\mathbf{\hat{h}_{1}}=\pm1$, the commutator Eq. (\ref{comm}) vanishes. This means that near the touching band points $[H_1,H_0]\approx 0 $, so we can approximate the time evolution operator Eq. (\ref{eqUeff}) as
\begin{equation}
\begin{split}
\mathcal{U}(k_x,k_y,\tau) \approx \exp{\left\{-i\tau (H_1+H_0)+\tau^2 [H_1,H_0]/2\right\}}
\label{hserie}
\end{split}
\end{equation}
where we used the Baker-Campbell-Hausdorff formula keeping terms up to order $\tau^2$. The second step is to expand $\mathcal{U}(k_x,k_y,\tau)$ around the neighborhood of touching band points, {\it i.e.}, we calculate the Taylor series of $\mathcal{U}(k_x,k_y,\tau)$ around $k_x=k_x^{*}$ and $k_y=k_y^{*}$.

After some algebraic manipulations we obtain 
\begin{equation}
\mathcal{U}(q_x,q_y,\tau)\approx \exp{\left[-i h_{T}\,\mathbf{\hat{h}}_{T}\cdot\mathbf{\sigma}\right]}
\label{approxU}
\end{equation}
where
\begin{equation}
\mathbf{h}_{T}=A(\lambda,\tau)\,q_x\mathbf{\hat{e}_x}+B(\lambda,\tau)\,q_y\mathbf{\hat{e}_y}+C(\lambda,\tau)\,q_y\mathbf{\hat{e}_z},
\label{topHam}
\end{equation}
with $q_x=k_x-k_x^{*}$, $q_y=k_y-k_y^{*}$, $\mathbf{\hat{h}}_{T}=\mathbf{h}_{T}/h_T$, $h_T=\left|\mathbf{h}_{T}\right|$, and
\begin{equation}
\begin{split}
A(\lambda,\tau)&=n\pi+\sqrt{3}(\lambda-1)\tau\sqrt{1+\frac{\left(1+\lambda/2-n\pi/\tau\right)^2}{4(\lambda-1)^2}}\\
B(\lambda,\tau)&=\frac{3}{4}(2+\lambda)\tau\\
C(\lambda,\tau)&=\frac{9\lambda\tau[(2+\lambda)\tau-2n\pi]}{8(\lambda-1)}.
\end{split}
\end{equation}
The topological properties of the system around the touching band points are given by the approximated effective Hamiltonian $\mathbf{\hat{h}}_{T}\cdot\mathbf{\sigma}$. To see that, 
note that near the touching band points $h_T\approx \pm n\pi$, the time evolution operator Eq. (\ref{approxU}) can be expanded as
\begin{equation}
\begin{split}
\mathcal{U}(q_x,q_y,\tau)&= \cos{(h_T)}-i(\mathbf{\hat{h}}_{T}\cdot\mathbf{\sigma})\sin{h_T}\\
&\approx \mathbf{1} -h_T(\mathbf{\hat{h}}_{T}\cdot\mathbf{\sigma}).
\end{split}
\end{equation}
Hence, all the topological features of the system will be given by $(\mathbf{\hat{h}}_{T}\cdot\mathbf{\sigma})$. The third step is to find the eigenvectors of $(\mathbf{\hat{h}}_{T}\cdot\mathbf{\sigma})$. It can be proven that they are given by the following spinors 
\begin{equation}
\begin{split}
\ket{\psi_{q'}^{\uparrow}}&= \frac{1}{\sqrt{2}}\left( \begin{array}{ccc}
\sqrt{1+\frac{C\,q{'}_y}{B\,h_T}} \\
e^{i\xi\alpha_{q{'}}}\sqrt{1-\frac{C\,q{'}_y}{B\,h_T}} \end{array} \right)\\
\ket{\psi_{q{'}}^{\downarrow}}&= -\frac{1}{\sqrt{2}}\left( \begin{array}{ccc}
e^{-i\xi\alpha_{q{'}}}\sqrt{1-\frac{C\,q{'}_y}{B\,h_T}} \\
 -\sqrt{1+\frac{C\,q{'}_y}{B\,h_T}}\end{array} \right)
\end{split}
\end{equation}
where $\xi$ can take the values $\xi=+1$ which corresponds to $+k_x^{*}$ and $\xi=-1$ to $-k_x^{*}$. We have used a new set of variables defined by
\begin{equation}
\begin{split}
q{'}_x&=q_x/A\\
q{'}_y&=q_y/B.
\end{split}
\end{equation}
and $\alpha_{q^{\prime}}$ is given by,
\begin{equation}
\alpha_{q{'}}=\tan^{-1}{\left(\frac{q{'}_y}{q{'}_x}\right)}.
\end{equation}

The four step is to compute the Berry phase directly from the definition, Eq. (\ref{Berry}). We start by calculating the Berry connection for $\xi=1$. We obtain that,
\begin{equation}
\mathbf{A}=\frac{1}{2}\left(1-\frac{C}{B\,h_{T}}q{'}_y\right)\nabla_{q{'}}\alpha_{q{'}},
\end{equation}
where 
\begin{equation}
\nabla_{q{'}}\alpha_{q{'}}=\frac{-q{'}_y\,\mathbf{\hat{e}}_x+q{'}_x\,\mathbf{\hat{e}}_y}{(q{'}_x)^2+(q{'}_y)^2}.
\end{equation}
Finally, we just calculate the Berry phase along a circumference centered at $q{'}_x=q{'}_y=0$. By using polar coordinates, $q{'}_x=q{'}\cos{\theta}$ and $q{'}_y=q{'}\sin{\theta}$ where $(q{'})^2=(q{'}_x)^2+(q{'}_y)^2$, we obtain 
\begin{equation}
\begin{split}
\gamma_C&=\int_{0}^{2\pi}\,\mathbf{A}\cdot d\mathbf{q{'}}\\
&=\frac{1}{2}\int_{0}^{2\pi}\left(1-\frac{\frac{C}{B}\sin{\theta}}{\sqrt{1+\frac{C^2}{B^2}\sin^2{\theta}}}\right)d\theta=\pi.
\end{split}
\end{equation}

A similar calculation can be done for $\xi=-1$, which gives $\gamma_{C}=-\pi$. Now the origin of the flat bands is clear, as they have a similar origin as for flat bands on Weyl semimetals, {\it i.e.} they are Fermi arcs which join two inequivalent Dirac cones with opposite Berry phase. However, for the special cases of  resonant driving $\tau=n\tau_c^{+}$, there is always one touching point at $k_x^{*}=0$ and $k_y^{*}=0,\,\pm2\pi/3$.  It
has $0$ or $\pm\pi$ quasienergy depending on $n$ (with $n\neq0$). At this point, the Berry phase is equal to zero. If we increase $\tau$ by a small amount, such point splits in two touching band points with opposite Berry phase. Hence, if the considered system is finite, an edge state joining such points will emerge, as it happens in pristine graphene nanoribbons or in Weyl semimetals. For the particular case $n=0$, touching band points are the same as in the time-independent case, thus their topological properties are the same as in zigzag graphene nanoribbons, namely, a flat band joining two inequivalent Dirac cones with opposite Berry phase emerges\cite{Volovik2011,torres2014}. Although the commutator Eq. (\ref{comm}) is zero at the touching band points studied here, away from such points the commutator Eq. (\ref{comm}) is no longer zero but proportional to $\sigma_{z}$, in other words, a mass-like term appears and a gap between touching band points is open. 

Finally, the range where edge states will emerge can be inferred from Eqs. (\ref{omegapm}) and (\ref{kx}), for $n=0$ is given by $|k_x|\geq k_x^{*(-)}$. For edge states with $n\neq0$, the interval where they appear in momentum space is given by the intersection of the solutions of $|k_x|\leq k_x^{*(+)}$ and $|k_x|\leq k_x^{*(-)}$. Then, we can create touching band points just by increasing the period of the driving $\tau$. In the next section we will discuss the experimental feasibility of the model studied here.

\section{Experimental feasibility}
\label{experiment}

%
\begin{figure}
\includegraphics[scale=0.39]{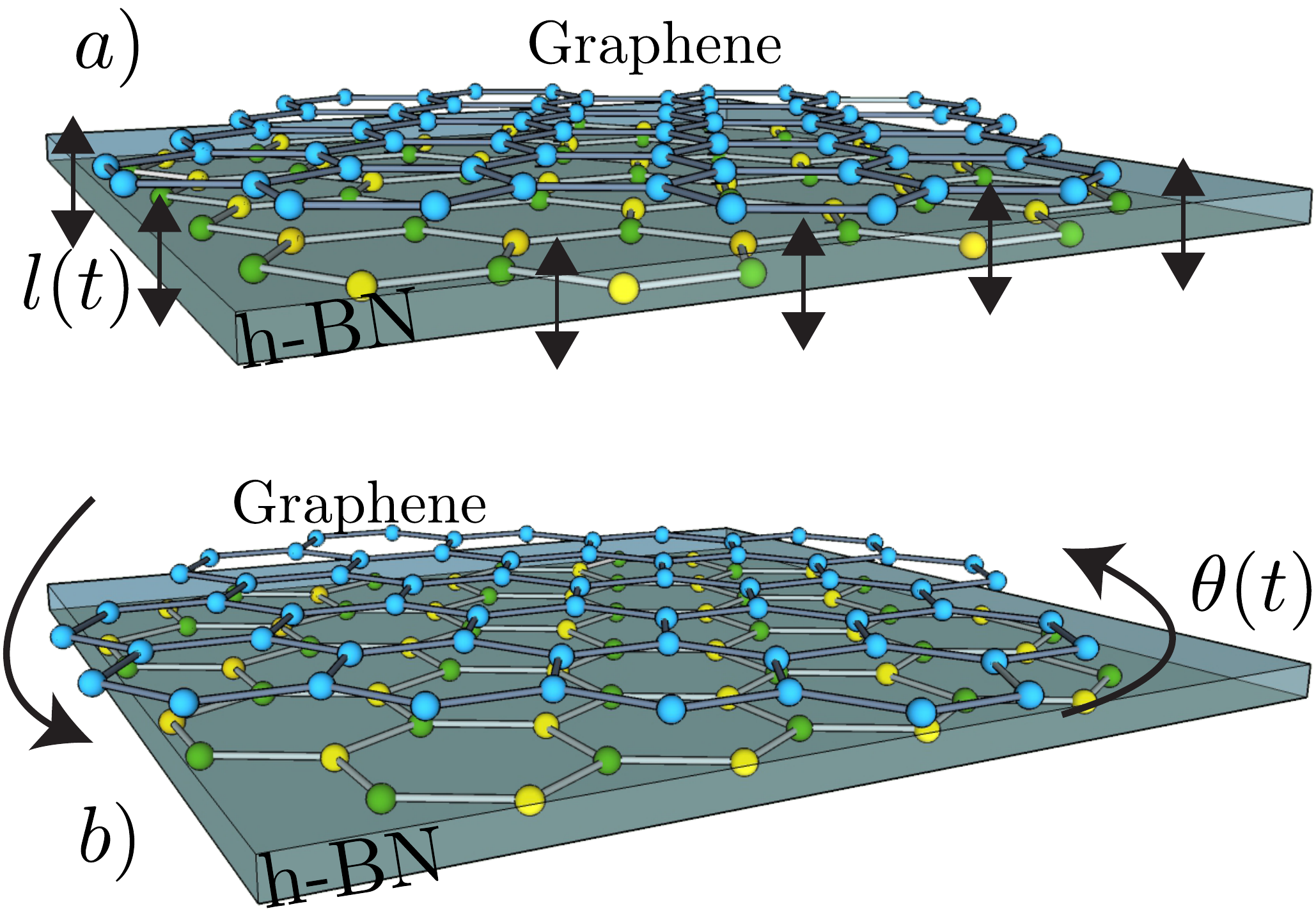}
\caption{(Color online.) Experiments proposed to observe topological flat bands in strained ZGN. As is shown, this can be achieved by placing a graphene monolayer over hexagonal boron nitride (h-BN). Then, the substrate can be moved up and down (a) or twisted (b) via a fast motor.}
\label{exp}
\end{figure} 

In this section we discuss the experimental feasibility of our model. We start by making a numerical estimation of the kicking frequency needed to observe the results obtained here. From Eq. (\ref{taumin}) the critical value of the driving period at which topological flat bands emerge is,
\begin{equation}
T=\frac{2\pi \hbar n}{3(2-\lambda)}.
\end{equation}
By introducing the numerical values, we obtain a driving period of $T\approx n\times 10^{-16} $ s. This kicking period is too small to be applied, however it grows with $n$, so for $n=10$ we have $T\approx 10^{-15}$ s. To observe this kind of effect, some experiments can be proposed. The first kind that one can imagine is to apply a time-dependent stress at the boundaries of the graphene membrane. Unfortunately, this experiment will not be able to discern the proposed effects, since stress is transmitted within graphene by phonons, which have a frequency very close to the proposed kicking frequency. This kind of experiment does not exhaust the options. We propose two different kinds of experiments to achieve such driving period. They are shown in Fig. \ref{exp}, the first one, panel a), consists of a graphene monolayer above an hexagonal boron nitride (h-BN) substrate, the substrate can be moved up and down by using different kinds of fast devices. In Fig. \ref{exp} a), the distance between graphene and h-BN, denoted by $l(t)$, is time-dependent. Similarly, the h-BN can be periodically twisted by an angle $\theta(t)$, as is shown in Fig. \ref{exp} b). The advantages of these experiments is that the strain field is applied at the same time at all lattices sites, and thus phonons are not needed to produce the strain field.

On the other hand, the delta kicking can be hard to be experimentally realized. Let us consider a more realistic kind of driving: harmonic driving. In particular, we chose a cosine time modulation given by,
\begin{equation}
\begin{split}
&\gamma_j(t)=\gamma_0\\
&+\cos{\left(\Omega t\right)}\gamma_0\lambda \xi(j+1) \sin\left[\pi \sigma \xi(j)\right] \sin(2 \pi \sigma j+\phi).
\label{gamcosine}
\end{split}
\end{equation}
Then, we can write the time-dependent Hamiltonian of the system as 
\begin{equation}
H(t)=H_0+\cos{\left(\Omega t\right)}H_1,
\label{cosineH}
\end{equation} 
where
\begin{equation}
\begin{split}
H_0&=\gamma_{0}\sum^{N-1}_{j=1} \left[a_{2j+1}^{\dag} b_{2j}+c(k_x)\,a_{2j-1}^{\dag} b_{2j}\right]+\mathrm{h.c.}\\
H_1&=\sum^{N-1}_{j=1}\left[\delta\gamma_{2j} a_{2j+1}^{\dag} b_{2j}+c(k_x)\, \delta\gamma_{2j-1} a_{2j-1}^{\dag} b_{2j}\right]\\
 &+\mathrm{h.c.}
\end{split}
\end{equation}
where $\delta\gamma_{j}=\gamma_{j}-\gamma_{0}$, see Eq. (\ref{gammas}). Since $H(t+T)=H(t)$ (here $T=2\pi/\Omega$), the Floquet theorem indicates that the wave functions of $H(t)$ can be written in terms of the fundamental frequency $\Omega$ as
\begin{equation}
\ket{\psi_{n\,j}(\mathbf{k},t)}=e^{-i\epsilon_n(\mathbf{k})\, t/\hbar}\sum_{m=-\infty}^{\infty}\ket{\varphi_{n,j}^{(m)}}e^{im\Omega t},
\label{kettime}
\end{equation}
where the coefficients $\ket{\varphi_{n,j}^{(m)}}$ at site $j$ satisfy the time-independent Schr\"odinger equation\cite{Rudner13},
\begin{equation}
\sum_{j^{\prime},m^{\prime}}\mathcal{H}_{j,j^{\prime}}^{m,m^{\prime}}\ket{\varphi_{n,j^{\prime}}^{(m^{\prime})}}=\epsilon_n\ket{\varphi_{n,j}^{(m)}},
\label{sequation}
\end{equation}
where $\mathcal{H}$, called the Floquet Hamiltonian, is given by,
\begin{equation}
\mathcal{H}_{j,j^{\prime}}^{m,m^{\prime}}=m\Omega\,\delta_{m,m^{\prime}}+\frac{1}{T}\int_{0}^{T}e^{-i(m-m^{\prime})\Omega t}H(t)\,dt.
\label{fham}
\end{equation}
Note that Eq. (\ref{sequation}) has solutions for each value of $\mathbf{k}$ all over $-\infty\leq\epsilon_{n}\leq\infty$. For our purposes, it is enough to consider just the first Brillouin zone of the Floquet space, {\it i.e.} $-\pi\leq\tau\epsilon_{n}\leq\pi$, with $\tau=T/\hbar$. 

\begin{figure}
\includegraphics[scale=0.41]{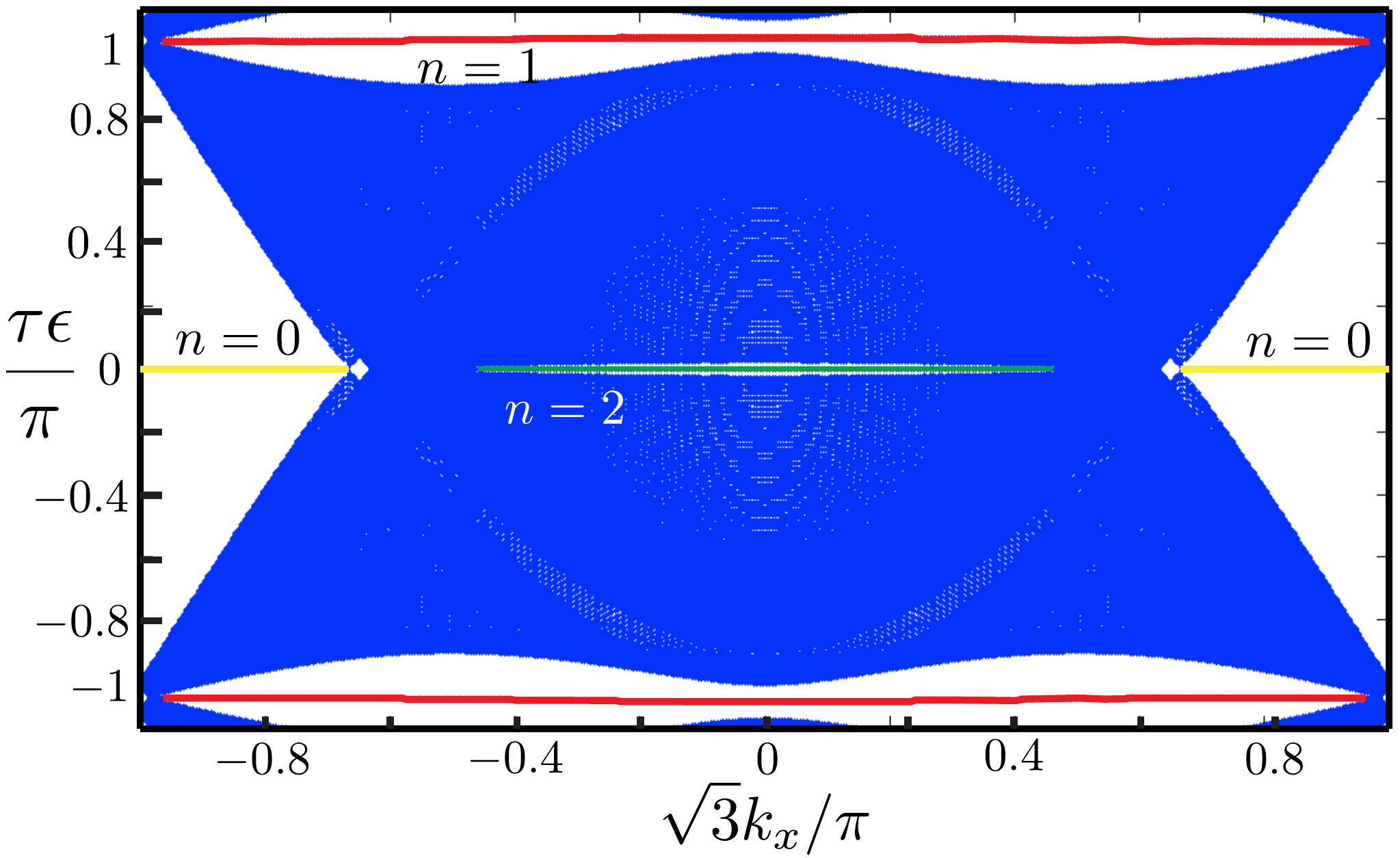}
\caption{(Color online.) Quasienergy spectrum obtained from Eq. (\ref{fham}) using $\tau=3$, $\lambda=0.1$, $\sigma=1/2$, $\phi=4\pi\sigma/3$, $N=240$ using fixed boundary conditions. Edge states for $n=0$ (time-independent edge modes at zero quasienergy) are indicated by solid yellow lines, whereas time-dependent edge states for $\tau\epsilon=\pm\pi$ ($\tau\epsilon=0$) are represented by solid red (green) lines. Note that the gaps separating time dependent edge states are smaller than the ones obtained by using a delta kicking, see Fig. \ref{bszt}. In addition, edge states, at $\pm\pi$ quasienergy, are no longer flat bands but dipersive edge modes.}
\label{cosine}
\end{figure} 

For a Hamiltonian given by Eq. (\ref{cosineH}), the Floquet Hamiltonian, Eq. (\ref{fham}), has a block trigonal form\cite{Rudner13}, where each block is a $N\times N$ matrix. As a first approximation, the quasienergy spectrum is well described by considering\cite{Rudner13} $-1\leq m\leq 1$. In Fig. \ref{cosine}, we present the quasienergy spectrum of $\mathcal{H}_{j,j^{\prime}}^{m,m^{\prime}}$ for $-1\leq m\leq 1$, $\lambda=0.1$, $\tau=3$, $\sigma=1/2$, $\phi=4\pi\sigma/3$, and $N=240$, calculated using fixed boundary conditions. As can be seen, time-independent flat bands still emerge at zero quasienergy, but the time-dependent flat bands at zero quasienergy are almost within the bulk spectrum (see Fig. \ref{cosine}, where such states are indicated by solid green lines). However, edge states at the edges of the first Brillouin zone of the Floquet space are still emerging, but they are no longer flat bands, in fact they have a small curvature as can be seen in Fig. \ref{cosine}, where such edge states are labeled by solid red lines. From the numerical results it seems that the gap that separates edge states from the bulk tends to be reduced by introducing a cosine modulation. To clarify that point let us make a comparison between the gaps that separate flat bands from the bulk states for delta and harmonic driving. We chose edge states around $\pm\pi$ quasienergy since for these states there is a well defined gap. At $k_x=0$, the gap is $\Delta\approx0.1 \,\text{eV}$ for the delta-kicking and  $\Delta\approx0.05 \, \text{eV}$ for the harmonic driving. This means that the gap obtained for the delta kicking is twice the one obtained for cosine kicking. Therefore, for the harmonic driving, the experimental observation of edge states is harder.
Even in the worst scenery, where the experiments proposed cannot be achieved, artificial lattices are good candidates for the experimental realization of our model, since in such lattices the hopping parameters can be tuned at will\cite{soltan11,soltana11,jotzu14,Messer15,weinberg16,EffOliva16}. Also, there is a recent proposal to use light to induce strain in graphene \cite{Salary16}, which is in the order of the required time-deformation driving.

\section{Conclusions}
\label{conclusion}
We have found topological non-trivial flat bands in time periodically driven strained graphene within the Floquet approach and in the limit of small strain's amplitude. This result was obtained using analytical calculations and compared with numerical calculations. An excellent agreement was found between them. That flat bands were understood as a kind of Fermi arcs joining nodal points (points at which the quasienergy spectrum takes zero or $\pm\pi$ values). Such points were characterized and have found to posses opposite Berry phases, which explain the emergence of flat bands between them. Moreover, our model provides a very simple picture about the emergence of such kind of flat bands in more complicated models and gives a very simple way to count the number of flat bands. Additionally, the experimental feasibility of the model was discussed and a more realistic time perturbation was studied. We found that, in the presence of a more realistic sinusoidal time perturbation, the main results of the paper are not modified: we still found edge states at zero and $\pm\pi$ quasienergy, although they are no longer flat bands. In addition, the gap that separates edge states from bulk states is bigger when a delta kicking driving is applied. In fact, the gap for harmonic driving is reduced almost to a half of the gap observed in delta driving.

This project was supported by DGAPA-PAPIIT Project 102717. P. R.-T. acknowledges financial support from Consejo Nacional de Ciencia y Tecnolog\'ia (CONACYT) (M\'exico).

\appendix
\section{}

First of all, let us calculate the commutator between $H_1$ and $H_0$ given by Eq. (\ref{hams}). We have,
\begin{equation}
\begin{split}
\left[H_1,H_0\right]&=\left[h_1^{(x)}\sigma_x+h_1^{(y)}\sigma_y,h_0^{(x)}\sigma_x+h_0^{(y)}\sigma_y\right]\\
&=h_0^{(y)}h_1^{(x)}\left[\sigma_x,\sigma_y\right]+h_0^{(x)}h_1^{(y)}\left[\sigma_y,\sigma_x\right]\\
&=2i\left(h_0^{(y)}h_1^{(x)}-h_1^{(y)}h_0^{(x)}\right)\sigma_z\\
&=-6i\lambda\sin{\left(3k_y/2\right)}\cos{\left(\sqrt{3}k_x/2\right)}\sigma_z.
\end{split}
\label{commA}
\end{equation}

Even though $H_1$ and $H_0$ do not commute, we can write equation (\ref{uop}) as
\begin{equation}
U(k_x,k_y,\tau)=\exp{\left[-i\tau H_{\mathrm{eff}}(k_x,k_y)\right]}.
\end{equation}
To do that we will use the addition rule of SU(3), namely,
\begin{equation}
e^{ia(\hat{n}\cdot\mathbf{\sigma})}e^{ib(\hat{m}\cdot\mathbf{\sigma})}=e^{-ic(\hat{g}\cdot\mathbf{\sigma})}
\label{addrule}
\end{equation}
here
\begin{equation}
\cos{c}=\cos{a}\cos{b}-\hat{n}\cdot\hat{m}\sin{a}\sin{b}
\label{energy}
\end{equation}
and
\begin{equation}
\hat{g}=\frac{1}{\sin{c}}(\hat{n}\sin{a}\cos{b}+\hat{m}\sin{b}\cos{a}-\hat{n}\times\hat{m}\sin{a}\sin{b}).
\label{unit}
\end{equation}

In our case we have that the Hamiltonians $H_1$ and $H_0$ can be written as 
\begin{equation}
\begin{split}
H_0(k_x,k_y)&=h_0(k_x,k_y)\hat{\mathbf{h}}_0\cdot\mathbf{\sigma}\\
H_1(k_x,k_y)&=h_1(k_x,k_y)\hat{\mathbf{h}}_1\cdot\mathbf{\sigma}
\end{split}
\end{equation}
where 
\begin{equation}
\begin{split}
\hat{\mathbf{h}}_0&=\frac{1}{h_0}\left(h_0^{(x)}(k_x,k_y)\hat{e}_x+h_0^{(y)}(k_x,k_y)\hat{e}_y\right)\\
\hat{\mathbf{h}}_1&=\frac{1}{h_1}\left(h_1^{(x)}(k_x,k_y)\hat{e}_x+h_1^{(y)}(k_x,k_y)\hat{e}_y\right)
\end{split}
\end{equation}
and
\begin{equation}
\begin{split}
h_0&=\left |\mathbf{h_0}(k_x,k_y)\right |=\sqrt{\left(h_0^{(x)}\right)^2+\left(h_0^{(y)}\right)^2}\\
h_1&=\left |\mathbf{h_1}(k_x,k_y)\right |=\sqrt{\left(h_1^{(x)}\right)^2+\left(h_1^{(y)}\right)^2}
\end{split}
\end{equation}
where we have not written the explicit dependence on $k_x$ and $k_y$ of $h_0,\, h_1,\, \mathbf{h_0}$, and $\mathbf{h_1}$ for the sake of simplicity.

Now, using the last part of equation (\ref{addrule}), the time evolution operator Eq. (\ref{uop}) takes the following form
\begin{equation}
U(k_x,k_y,\tau)=e^{-ia\tau(\hat{\mathbf{h}}_1\cdot\mathbf{\sigma})}e^{-ib\tau(\hat{\mathbf{h}}_1\cdot\mathbf{\sigma})}=e^{-i\omega\tau(\mathbf{\hat{h}_{\mathrm{eff}}}\cdot\mathbf{\sigma})}.
\end{equation}
As we can see, by using the addition rule of SU(2) the time evolution operator is diagonalized. The quasienergies can be obtained from Eq. (\ref{energy}) and are given by
\begin{equation}
\begin{split}
&\cos{\left[\tau\omega(k_x,k_y)\right]}=\cos{(\tau h_0)}\cos{(\tau h_1)}-\\
&\hat{\mathbf{h}}_1\cdot\hat{\mathbf{h}}_0\sin{(\tau h_0)}\sin{(\tau h_1)}
\end{split}
\end{equation}
where
\begin{equation}
\begin{split}
\hat{\mathbf{h}}_1\cdot\hat{\mathbf{h}}_0&=\frac{\lambda}{h_0h_1}\left[-4\cos^2{\left(\sqrt{3}k_x/2\right)}\right]\\
&\frac{\lambda}{h_0h_1}\left[-\cos{\left(\sqrt{3}k_x/2\right)}\cos{\left(\frac{3k_y}{2}\right)}+\frac{1}{2}\right]
\end{split}
\label{dvec}
\end{equation}
The unit vector $\mathbf{\hat{h}_{\mathrm{eff}}}$ can be obtained from Eq. (\ref{unit}), we have
\begin{equation}
\begin{split}
\mathbf{\hat{h}_{\mathrm{eff}}}&=-\frac{1}{\sin{\left(\tau\omega\right)}}\left[\hat{\mathbf{h}}_1\sin{(\tau h_1)}\cos{(\tau h_0)}\right]\\
&-\frac{1}{\sin{\left(\tau\omega\right)}}\left[\hat{\mathbf{h}}_0\sin{(\tau h_0)}\cos{(\tau h_1)}\right]\\
&-\frac{1}{\sin{\left(\tau\omega\right)}}\left[\hat{\mathbf{h}}_1\times\hat{\mathbf{h}}_0\sin{(\tau h_1)}\sin{(\tau h_0)}\right]
\end{split}
\end{equation}
with
\begin{equation}
\hat{\mathbf{h}}_1\times\hat{\mathbf{h}}_0=\frac{3\lambda}{h_0h_1}\left[\sin{\left(3k_y/2\right)}\cos{\left(\sqrt{3}k_x/2\right)}\right]\,\hat{e}_z.
\label{pvec}
\end{equation}
Finally, the effective Hamiltonian is
\begin{equation}
H_{\mathrm{eff}}(k_x,k_y)=\omega(k_x,k_y)\,\mathbf{\hat{h}_{\mathrm{eff}}}\cdot\mathbf{\sigma}.
\end{equation}

\bibliography{biblioArmChairGraphene}{}
\end{document}